# Paddle-wheels, friction, and moving-boundary work

Andrea Crespi

Dipartimento di Fisica - Politecnico di Milano, piazza Leonardo da Vinci 32, 20133 Milano, Italy

E-mail: andrea.crespi@polimi.it

## Abstract

Moving-boundary work, also called pressure-volume work or expansion work, is perhaps the main form of work discussed in introductory courses about thermodynamics. Here, we take a particular definition of this kind of thermodynamic work. On this basis, we show with simple reasoning that significant phenomena involving dissipation of mechanical energy, often mentioned as well in those courses but with a vague formalization, can be traced back right to moving-boundary processes. We refer in particular to the conversion of external work into internal energy in paddle-wheel experiments such as the one conducted by J. P. Joule in 1850, and to other processes involving friction.

## 1. Introduction

The concept of work is ubiquitous in thermodynamics, indeed as much as the one of heat. Nevertheless, the rigorous definition of thermodynamic work, and even its role in the theory (whether or not it should be assumed as a fundamental concept) has been object of an enduring debate [1–7].

Thermodynamics teaching, on the other hand, especially at the introductory level, is traditionally focused on the study of ideal gases or fluids. In this context, across the variety of possible approaches and textbooks [8–16], emphasis is generally given to the definition of moving-boundary work, called also pressure-volume work or expansion work, according to

$$\mathcal{W} = \int_i^f p \, dV \tag{1}$$

where $i$ and $f$ indicate the initial and final states of the transformation, $V$ is the volume of the system, and $p$ is a pressure which may be referred to the system or the surroundings, depending on the approach adopted [5]. A minus sign may be added in front of the integral, in agreement with the chosen sign conventions.

In addition to this, textbooks may provide a heterogeneous list of additional expressions for the thermodynamic work to be employed for other types of systems or in different situations (see e.g. [9] or [12]). Such expressions are justified from global considerations on the physical processes occurring to the system or to the surroundings, and they incorporate notions from mechanics or electromagnetism whenever it is needed. From the educational point of view, the naturalness by which the most diverse forms of work are employed in the framework of the first principle of thermodynamics is certainly valuable, in order to illustrate the ubiquity of energy conservation. On the other hand, the details on how the thermodynamic work is actually exchanged between the system and its surroundings may get lost or remain hidden.

An emblematic case is the didactic discussion of the so-called paddle-wheel experiment, by which James Prescott Joule first quantified the mechanical equivalent of heat [17]. This experiment is of paramount importance, both in the historical development of physics [18, 19] and in physics teaching, to illustrate the concept of energy [20, 21]. In Joule's apparatus, a certain quantity of water is contained in an adiabatic calorimeter, and constitutes the thermodynamic system of interest. The calorimeter is equipped with a paddle-wheel mounted on a rotating shaft. The experiment consists in activating the rotation of the paddle-wheel, immersed in water, by means of falling weights. The latter are connected to the shaft by a mechanical system of ropes and pulleys. The crucial observation is that a temperature increase of the water, which could be equivalently produced by heat transfer, is associated proportionally to the mechanical work exerted on the system.

Quantification of the mechanical work is typically given from considerations about the mechanical energy of the falling weights (which, anyway, should be carried out correctly [21]). How this work is transferred effectively to the water in the calorimeter remains vague. Yet, the water included in the calorimeter is a homogeneous fluid, namely an elementary enough thermodynamic system that should be accessible to more detailed analysis.

Here, we recall a sufficiently general definition for the moving-boundary thermodynamic work and, by leveraging this definition, we discuss with simple considerations how one can interpret the work exerted on a system in a few cases of interest. These include the above-mentioned paddle-wheel experiment, the deformation process of a solid with consequent heating, and dry friction. Note that our purpose is not to calculate effectively the work contribution with (1) by means of data that could be considered as known, in reasonable experimental situations. We desire instead to show how, in these relevant examples, the exchange of work between the system and the surroundings can be interpreted as boundary work, whereas it might not be usually depicted in this way.

## 2. Simple systems

A thermodynamic system is a collection of particles or bodies enclosed in a defined boundary surface $S$. Homogeneous substances consisting of a single chemical phase, and with a definite chemical composition, are usually modelled as hydrostatic systems, characterized by three coordinates such as pressure $p$, volume $V$, and temperature $T$ [8, 9, 11, 12].

In the absence of the exchange of matter with the surroundings (namely, if the system is closed), only two of the three coordinates are independent and are sufficient to identify a given equilibrium state, as well as to determine the values of the thermodynamic state functions. The internal energy of hydrostatic systems may thus be written in alternative forms as

$$U = U(p, V) = U(p, T) = U(T, V). \qquad (2)$$

One notes that in these systems only one coordinate (the total volume) is actually being adopted to describe the external shape of the boundary surface. This sounds like a good description, for instance, for a gas in a cylinder enclosed by a mobile piston, or for a liquid contained in a given rigid vessel. In these cases, the specification of the total volume allows in principle, without ambiguities, determination of the shape of the boundary surface.

The definition of thermodynamic work that will be discussed in the next section is conveniently introduced in the framework of slightly more involved systems, whose boundary surface may need a description in terms of more than one coordinate. Namely, we allow the boundary surface not only to change the total enclosed volume $V$, but also to change its shape according to a number of deformation parameters. These systems may be called, following Carathéodory [22], *simple* systems.

Such deformation parameters could be taken as dimensionless quantities or, multiplied by proper constants, as homogeneous to volumes. An example of a simple system according to the above definition could be a certain quantity of gas contained in a special vessel that is connected to $n$ cylinders, each closed with a mobile piston. The shape of the boundary surface is defined univocally once the volumes of the various cylinders are given, or equivalently, once the total volume $V$ is given, together with the volumes $V_1, \ldots, V_{n-1}$ of $n-1$ of these cylinders. The internal energy of the system may be written as

$$U = U(T, V, V_1, V_2, \ldots, V_{n-1}). \qquad (3)$$

If a slightly different set of independent coordinates is chosen, taking the pressure $p$ instead of the overall volume $V$, we may also write

$$U = U(T, p, V_1, V_2, \ldots, V_{n-1}). \qquad (4)$$

At the limit of increasing complexity, the set of parameters that describe the shape of the surface becomes a mathematical function that represents in full detail the three-dimensional surface $S$ of the boundary.

## 3. Moving-boundary work

In classical thermodynamics one typically neglects long-range interactions between the system and the surroundings, such as gravitational or electromagnetic ones. Hence, it is assumed that all relevant interactions occur across the boundary and by means of direct-contact forces. We shall further neglect shears and only consider pressure forces acting normally to the surface. In this scenario, we can reach our expression for the moving-boundary work with a reasoning akin to the following.

Let us consider an infinitesimal element $dS$ of the boundary surface, and let us consider an infinitesimal thermodynamic transformation that displaces this element by $\vec{dr}$. Pressure forces act on the element of the surface both from inside and from outside, namely forces originate both from the bodies composing the system and from the bodies composing the

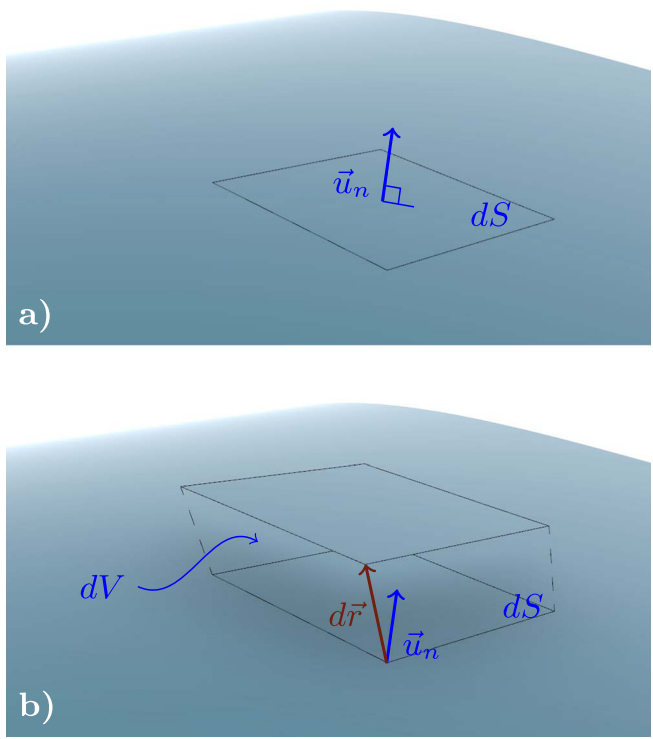

**Figure 1.** (a) An element *dS* of the boundary surface is associated to its normal unit vector $\vec{u}_n$, pointing outwards. (b) A displacement $d\vec{r}$ of the surface element corresponds to an increment $dV = (d\vec{r} \cdot \vec{u}_n)dS$ of the volume included in the boundary.

surroundings (figure 1). We may write

$$d\vec{F}_{\text{sys}} = p_{\text{in}} \, dS \, \vec{u}_n \tag{5}$$

$$d\vec{F}_{\text{sur}} = -p_{\text{ext}} \, dS \, \vec{u}_n \tag{6}$$

where $d\vec{F}_{\text{sys}}$ and $d\vec{F}_{\text{sur}}$ are, respectively, the infinitesimal forces acting on *dS* from inside and from outside, with $\vec{u}_n$ the unit vector orthogonal to *dS* pointing outwards. It is convenient to consider the pressures $p_{\text{in}}$ and $p_{\text{ext}}$ in the above equations just as a way to express the resultant forces (from inside or from outside) exerted on that element of area, without implying that they are uniform in larger regions. Namely, it is not assumed for the moment that these pressures are intensive coordinates characterizing the entirety or a component of the system or of the surroundings.

The boundary is an ideal surface with negligible mass. Whatever the acceleration of *dS*, the overall resultant of the applied forces, from inside and from outside, must remain null. The overall work, which is the sum of the work $\delta\mathcal{W}_{\text{sys}}$ of the forces applied to the boundary from the system, and of the work $\delta\mathcal{W}_{\text{sur}}$ applied to the boundary from the surroundings, also remains null.

$$\delta\mathcal{W}_{\text{tot}} = \delta\mathcal{W}_{\text{sur}} + \delta\mathcal{W}_{\text{sys}} = 0 \tag{7}$$

$$\delta\mathcal{W}_{\text{sys}} = -\delta\mathcal{W}_{\text{sur}} \tag{8}$$

From this point forward, one could follow different routes, focusing more on the system rather than on the surroundings [5]. Here, we choose to rely on the external pressure $p_{\text{ext}}$, as for instance in [11], and write

$$\delta\mathcal{W}_{\text{sys}} = -\delta\mathcal{W}_{\text{sur}} = -(\vec{dF}_{\text{sur}} \cdot \vec{dr}) = p_{\text{ext}} dS(\vec{dr} \cdot \vec{u}_n) = p_{\text{ext}} dV \tag{9}$$

where $dV = dS(\vec{dr} \cdot \vec{u}_n)$ corresponds indeed to the local infinitesimal variation of the volume *of the system* that is being caused by the infinitesimal displacement of the element of the surface $dS$.

In case the shape of the boundary is described by discrete volume-like coordinates, the infinitesimal work would be given by

$$\delta\mathcal{W} = \sum_i p_{\text{ext},i}\, dV_i \tag{10}$$

in agreement with the treatment in [22].

More generally, the overall work is calculated by integrating $\delta\mathcal{W}_{\text{sys}} = p_{\text{ext}} dV$ on two domains, namely both on the whole boundary surface and on the whole transformation:

$$\mathcal{W} = \mathcal{W}_{\text{sys}} = \int_{S, i \to f} p_{\text{ext}}\, dV \tag{11}$$

where $i$ and $f$ indicate the initial and final equilibrium states. Note that in expression (11), $dV$ does not indicate the differential of the total volume of the system but the local volume variation induced by the local displacement of the boundary as in figure 1. In addition, the external pressure $p_{\text{ext}}$ may vary at different points of the boundary surface. This means that non-vanishing work contributions may also result from two opposite local displacements of portions of the boundary, which by themselves produce a null variation of the overall system volume, if they happen against two different external pressures. We shall see indeed an example of this occurrence in the next section.

## 4. Paddles moving in a fluid

We discuss now how the work done by a paddle, translating while immersed in a fluid, may be interpreted as moving-boundary work in agreement with (11). The fluid shall be considered as the thermodynamic system, and the paddle-wheel shall be part of the surroundings.

We deal at first with a more intuitive configuration and consider a single paddle, partially immersed in the fluid. If we draw the boundary surface of the fluid, we obtain something similar to figure 2: the plane surface of the fluid presents a trench, in correspondence to the immersed part of the paddle. The paddle motion results in a translation of the trench, e.g. in the positive direction of the $x$ axis. The paddle acts on the boundary with forces that can be described in the terms of external pressures $p_{\text{ext}}$, which we allow to be different on each facet of the trench.

The lateral and bottom facets move along a direction orthogonal to their respective $\vec{u}_n$ vectors; therefore, pressure forces do not do work on these pieces of surface, according to the definition (11). Instead, the large front and back facets of area $\mathcal{A}$ translate in a direction parallel or antiparallel to their $\vec{u}_n$ vectors. On the one hand, facet ABCD has its normal vector coincident with $+\vec{u}_x$, and its infinitesimal displacement $\vec{dr} = dx\vec{u}_x$ produces a variation $dV = \mathcal{A}dx$ in the volume of the system. On the other hand, the normal vector to the facet

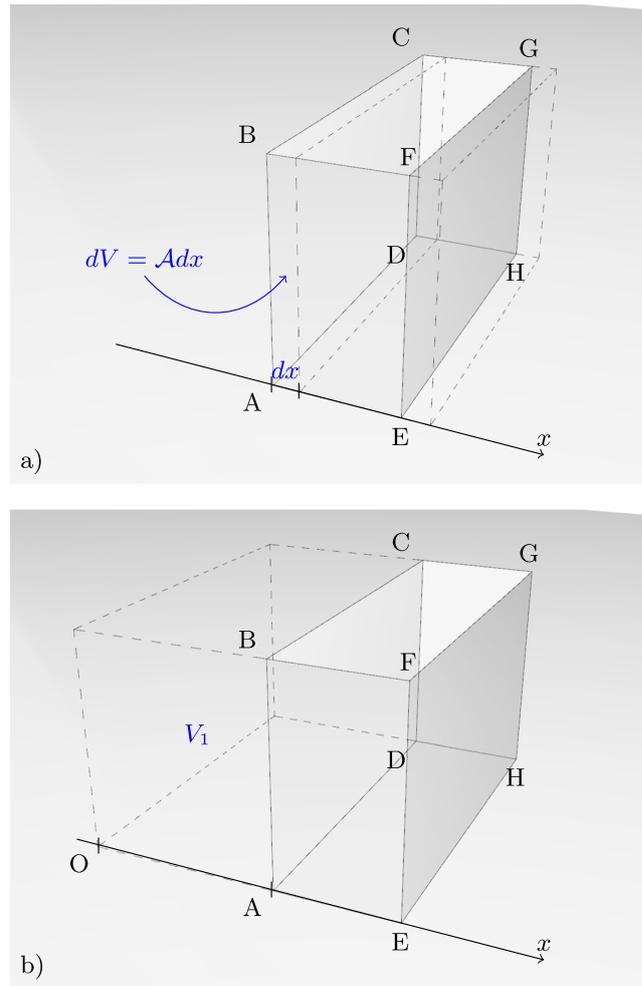

**Figure 2.** (a) A paddle immersed in a fluid deforms the boundary of the system: the plane top surface displays a trench. Translation of the paddle along the *x* axis (orthogonal to the largest facets of the trench, which have an area $\mathcal{A}$) causes a continuous deformation of the boundary surface and of the enclosed volume. An infinitesimal displacement *dx* corresponds to a potential increase of volume $dV = \mathcal{A}\,dx$ produced by the translation of the facet ABCD, and an equal decrease of volume $-dV = -\mathcal{A}\,dx$ produced by the parallel translation of the facet EFGH. (b) The position of the trench in the space can be specified by means of a volume-like coordinate $V_1$, which is the volume spanned by the surface ABCD in a translation along the *x* axis, starting from a reference origin O.

EFGH is $-\vec{u}_x$ and an infinitesimal displacement $d\vec{r} = -dx\,\vec{u}_x$ produces a volume variation $-dV = -\mathcal{A}dx$. (Note that the two variations sum algebraically to zero, and indeed the overall volume of the fluid is not changed.)

As mentioned, we allow the external pressure (i.e. the pressure exerted by the paddle) to be different on the facets ABCD and EFGH, and we indicate the two values respectively with $p_{\text{ext},1}$ and $p_{\text{ext},2}$. The motion of the paddle in the fluid realistically correspond to a case for which $p_{\text{ext},2} > p_{\text{ext},1}$, as the paddle is pushing the fluid forward, due to the effects of viscous friction. The resultant work for an infinitesimal displacement $dx$, calculated by integrating (9) on the whole boundary, is

$$\delta \mathcal{W}_{\text{paddle}} = (p_{\text{ext},1} - p_{\text{ext},2}) \, \mathcal{A} \, dx. \tag{12}$$

If $p_{ext,2} > p_{ext,1}$ the work is negative and we write, for the work done on the whole movement:

$$\mathcal{W}_{\text{paddle}} < 0. \tag{13}$$

The above analysis shows in an easy fashion that mechanical work is done on the fluid system by local displacements of portions of its boundary, under the action of pressure forces different in magnitude. In fact, if $p_{\text{ext},1} = p_{\text{ext},2}$, the resultant work would be vanishing.

Several different coordinates could be chosen to specify the position of the trench along the $x$ axis. For instance, we could specify the $x$ coordinate of the point A or of the point E. We may also adopt a volume-like coordinate $V_1$, consisting in the volume spanned by the facet ABCD along its translation, starting from a reference O fixed on the $x$ axis. Together with the volume $V$, this further coordinate $V_1$ fully determines the shape of the boundary surface. It is not difficult to note that the infinitesimal volume variation $dV = \mathcal{A} \, dx$ coincides with an infinitesimal variation of this coordinate:

$$dV_1 \equiv dV = \mathcal{A} \, dx. \tag{14}$$

Hence, the infinitesimal paddle work can be written as $\delta \mathcal{W}_{\text{paddle}} = (p_{\text{ext},1} - p_{\text{ext},2}) \, dV_1$ and the overall work as

$$\mathcal{W}_{\text{paddle}} = (p_{\text{ext},1} - p_{\text{ext},2}) \Delta V_1 \tag{15}$$

where $\Delta V_1$ expresses the variation of the coordinate $V_1$ in the transformation.

As a matter of fact, $\mathcal{W}_{\text{paddle}}$ is not the only contribution of work that is present in this experimental situation. Indeed, the fluid system as a whole may also change its total volume $V$ because of thermal expansion against the external atmospheric pressure $p_{\text{atm}}$. This gives rise to a further contribution:

$$\mathcal{W}_{\text{expansion}} = p_{\text{atm}} \Delta V \tag{16}$$

and the total work exchanged by the system is

$$\mathcal{W} = \mathcal{W}_{\text{paddle}} + \mathcal{W}_{\text{expansion}}. \tag{17}$$

Note that both contributions are the result of moving-boundary work according to (11). In particular, $\mathcal{W}_{\text{paddle}}$ comes from the simultaneous displacements of the boundaries of the trench, which do not produce net variations of the total volume, as discussed above. $\mathcal{W}_{\text{expansion}}$ has not been examined in detail here, but we might imagine it as resulting from a slight raising of the top surface of the fluid in the calorimeter, occurring against the atmospheric pressure, due to the increase in temperature in the process.

To conclude our analysis of this experiment, we can relate the temperature increase of the water to the paddle work, proceeding as follows. First, we may write the energy balance of the process, according to the first principle of thermodynamics:

$$Q - \mathcal{W} = \Delta U \tag{18}$$

$$-\mathcal{W}_{\text{paddle}} = \Delta U + \mathcal{W}_{\text{expansion}} = \Delta U + p_{\text{atm}} \Delta V \tag{19}$$

where we have considered $Q = 0$ as no heat has been exchanged with the surroundings.

For a thermodynamic transformation occurring at constant external pressure, it is convenient to introduce the enthalpy $H = U + pV$, and recall that the heat capacity at constant

pressure reads [11]

$$C_p = \left| \frac{\partial H}{\partial T} \right|_p. \tag{20}$$

We note that, by exploiting the definition of enthalpy, (19) is simply rewritten as

$$-\mathcal{W}_{\text{paddle}} = \Delta H. \tag{21}$$

In principle, the internal energy and the enthalpy may be functions of $(T, V, V_1)$. However, it is reasonable to assume that $V_1$ has no influence on the energy of the system, so we can consider this system as simply hydrostatic for the purpose of the energetic analysis. Hence, by applying (20) and assuming $C_p$ as constant, we conclude

$$\Delta T = \frac{\Delta H}{C_p} = -\frac{(p_{\text{ext},1} - p_{\text{ext},2})\Delta V_1}{C_p} > 0. \tag{22}$$

This description is generalized straightforwardly to the case of more paddles, and can be extended to the case of paddles completely submerged by considering rectangular, closed cavities, which translate in the fluid. In case where the motion is not a pure translation but a rotation around a fixed axis, as for the paddle-wheel of Joule's experiment, the pressures will not be uniform on the facets of the cavity. Opportune averages need to be considered instead.

## 5. Deformation and friction

The formalism of moving-boundary work is naturally apt to cases in which one observes a net variation of the system volume because of external pressure forces. In the former section, we have analyzed one case in which local pressure forces do work by deformation, but this work is not mainly due to a change in the total volume (even if a change in the volume may eventually occur, correlated to an increase in temperature). There are many other similar examples. An interesting situation that we may try to model is the bending deformation of a thin solid mechanical structure; more specifically, let us consider a small cantilever, fixed to a larger base made of the same material, which is bent (rightwards) under the action of external forces as in figure 3. In particular, there shall be forces acting on the left surface, not balanced by forces on the right surface. The cantilever with its base is our thermodynamical system, and we analyze how work is exchanged between the system and the surroundings.

Thermodynamic analysis of solids may be difficult even when they are composed of a single homogeneous substance. In fact, throughout their volume, stress components may be non-isotropic and non-uniform even in equilibrium states [23, 24]. Here, we aim at an approximated and simplified treatment compatible with a possible 'classroom illustration' and we assume shear forces on the boundary are negligible. This means that our solid cantilever can be included in the category of simple systems, described by the temperature $T$ and by the shape of the external boundary. We can then imagine a certain distribution of pressure forces, applied in given proportions to the various points of the left surface of the cantilever, which add to the uniform pressure on the whole boundary due to the surrounding air.

This framework allows us to employ fruitfully our definition of work (11). In agreement with the meaning of the volume element $dV$ in the definition, in the case of an infinitesimal bending rightward, the deformation of the right surface produces a positive volume variation, while the deformation of the left surface produces a negative volume variation. We can introduce a deformation coordinate $V_1$ that indicates the bending state, defined as the volume spanned by the lateral surface of the cantilever, starting from the rest position up to the desired bent position (see figure 3). In this way, a uniform additional pressure $p_1$ on the left surface of the cantilever produces a work on the system precisely equal to

$$\mathcal{W}_{\text{bending}} = -p_1 V_1 < 0. \tag{23}$$

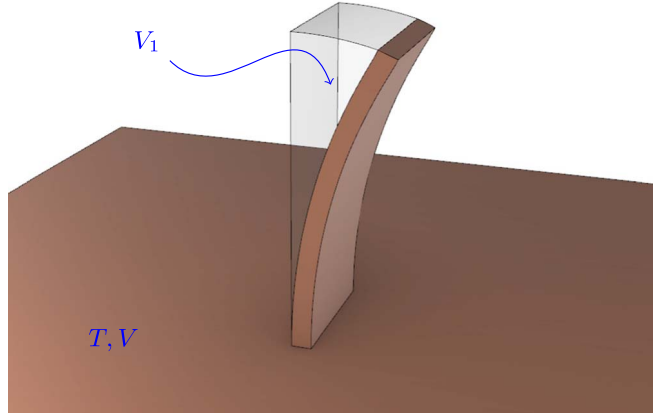

**Figure 3.** A cantilever fixed to a larger base, made of the same material, is modelled as a thermodynamic system characterized by three independent coordinates: the overall temperature $T$, the total volume $V$, and the deformation coordinate $V_1$ (homogeneous to a volume).

In the case of a non-uniform distribution of forces, with fixed proportions, we may get an analogous expression where $p_1$ is substituted by an average quantity judiciously defined.

The uniform atmospheric pressure acting on the whole boundary surface may do work as well, if the global volume has varied due to an increase in temperature, or because of 'side effects' of the deformation. As work is additive, we can consider this work contribution separately, as given by (16): $\mathcal{W}_{\text{expansion}} = p_{\text{atm}} \Delta V$.

If the process of deformation occurs adiabatically, because of the first principle of thermodynamics the net external work $\mathcal{W} = \mathcal{W}_{\text{bending}} + \mathcal{W}_{\text{expansion}}$ equals the variation of internal energy (changed in sign). The internal energy here is a function of at least three variables, such as $(T, V, V_1)$. Work may transform partially into elastic energy (a contribution that is mainly a function of $V_1$) and partially into thermal energy, i.e. into an increase in the temperature $T$. The proportions of this transformation depend non-trivially on the details of the process.

An experimental condition that is easily accessible to modelling may be the following. The cantilever with its base is kept in an environment with low pressure $p_0$, namely a reasonable degree of vacuum. Initially, forces are employed to bend the cantilever up to a certain flexion state. Then, the forces are released and the cantilever begins to oscillate. Oscillations are gradually damped by internal friction in the solid structure and, after some time (not necessarily short), the structure returns to the initial, undeformed, equilibrium state. Due to the vacuum environment, heat exchange with the surroundings is negligible. We can write the energy balance similarly to (18)–(19), and obtain analogously to (21):

$$-\mathcal{W}_{\text{bending}} = \Delta H. \tag{24}$$

We choose as as a set of variables $(T, p, V_1)$ instead of $(T, V, V_1)$, and write the enthalpy as

$$H = H(T, p_0, V_1) = U(T, p_0, V_1) - p_0 V(T, p_0, V_1) \tag{25}$$

where we have considered also that the overall volume may depend on $T$, $p$ and possibly $V_1$. The enthalpy variation for this whole process, in which the cantilever is undeformed both at the initial and final time, is

$$\Delta H = H(T_1, p_0, 0) - H(T_0, p_0, 0) \tag{26}$$

and is not affected by the deformation $V_1$. It is thus reasonable to write $\Delta H = C_p \Delta T$ and we conclude

$$\Delta T = \frac{H(T_1, p, 0) - H(T_0, p, 0)}{C_p} = -\frac{W_{\text{bending}}}{C_p}. \tag{27}$$

In the case where the air pressure is not negligible, the process can be discussed at least qualitatively. Oscillations of the cantilever may produce work on the surroundings, acting similarly to the paddles of the previous section (here swapping the roles of system and surroundings). Heat could also be exchanged. Both would definitely contribute to increase the temperature of the surroundings, up to the equilibrium point with the cantilever.

This example is significant, because simple models [25] for dry friction between sliding bodies may explain it via the continuous deformation, bending, and release of small dents, protruding from the surfaces. These phenomena occurring at the micrometric scale result macroscopically in the work of dissipative forces, from the point of view of mechanics, or equivalently in the transformation of external work into internal energy of the system with an increase in temperature, from the point of view of thermodynamics. The didactic presentation of this simple modeling of oscillation damping may serve as a hint to discuss qualitatively some features of dry friction processes.

## 6. Conclusions

We have shown that the concept of moving-boundary work, if suitably defined, can be useful to illustrate the system–surroundings interactions well beyond the elementary case of a gas contained in a cylinder closed by a mobile piston. Here, we have only gone through a few examples, as we have only discussed the work exchanged by water in a calorimeter, when it is stirred by paddles or similar tools, and the work of deformation of a solid object along well-determined modes. In fact, a *simple* thermodynamic system can exchange work only via displacements and deformations of its boundary [22]. Hence, (11) should be, in principle, more widely applicable to interpret the transformations of these systems.

It is obvious that (11) is not always practical to calculate the work contribution, because quantifying the distribution of pressure forces acting on the boundary, and the deformations that the latter undergoes, may be difficult. However, a qualitative identification of these pressure forces and of these deformations may provide useful insights on the mechanisms of the exchange of mechanical work in the analyzed thermodynamic transformations.

Illustrating simple examples, such as the ones treated here, might be valuable from an educational point of view, as they could contribute to strengthen the perceived connections between the different building blocks of thermodynamic theory. For instance, the same concept of work could be applied both in the initial definition of the quantity, and in the historical justification of the first principle of thermodynamics, with reference to Joule's paddle-wheel experiment.

**Data availability statement**

No new data were created or analyzed in this study.

**ORCID iDs**

Andrea Crespi  https://orcid.org/0000-0001-5352-6581